\documentclass{article}
\usepackage{amssymb}

\usepackage{amsmath}

\input{tcilatex}

\begin{document}

\begin{center}
{\Large First quantized electron and photon model of QED and radiative
processes} 

\bigskip 

\bigskip Melike B. YUCEL and Nuri UNAL

Akdeniz University, Physics Department, P. O. Box 510,

07058 Antalya, Turkey

Pacs: 03.75.-b, 11.10.-z, 12.20.-m, 13.40.-f, 14.60.Cd, 14.70.Bh, 32.80.Cy

\bigskip

\bigskip

\textbf{Abstract}
\end{center}

In this study we combine the classical models of the massive and massless
spinning particles, derive the current-current interaction Lagrangian of the
particles from the gauge transformations of the classical spinors, and
discuss radiative processes in electrodynamics by using the solutions of the
Dirac equation and the quantum wave equations of the photon. The
longitudinal polarized photon states give a new idea about the vacuum
concept in electrodynamics.

\bigskip

\section{Introduction}

\noindent \qquad The perturbative formulation of quantum electrodynamics is
based on at least four different approaches: (1) The standard operator
expansion in the quantum field theory $\left[ 1\right] $; (2) Green's
function expansion $\left[ 2\right] $; (3) S- matrix unitary expansion $%
\left[ 3\right] $ and (4) path integral quantization of classical particle
trajectories $\left[ 4\right] $. There are also various nonperturbative
approaches $\left[ 5,6\right] $. In some of these approaches electrons are
represented by the solution of the Dirac equation instead of the second
quantized Dirac fields. But photons are represented as second quantized
Maxweel fields, which have infinite degrees of freedom. In classical
electrodynamics the radiative processes and the radiation reaction are
described by the self-interactions of the electrons with themselves and also
in this formalism the electromagnetic field is represented by Green's
function.

Up to now there have been a lot of attempts to understand the dynamical
structure of the electron. The Heisenberg equations of the Dirac electron
give a helical trajectory as its natural free motion (the zitterbewegung).
Barut and Zanghi derived a classical analogue of the zitterbewegung
oscillations $\left[ 7\right] $, the quantization of this model gives the
Dirac equation as well as higher spin wave equations and the first quantized
formulation of the QED can be derived by the quantization of the classical
action of this model $\left[ 8\right] $.

Recently, one of us proposed a classical model for the massless spinning
particles and the quantization of this model gives a new wave equation for
the massless, spin-1 particles as well as the wave equation for Majarona
neutrinos $\left[ 9\right] $. \qquad \noindent\ \ \ \ 

The aim of this study is to derive a completely first quantized version of
the QED in which the electrons and photons are described classically as
spinning particles with a finite degree of freedom. In Section II, we
discuss the free particle Lagrangian of the massive and massless spinning
particles and try to derive the interaction term by using a new global gauge
principle for the classical spinors of these particles. The local form of
the gauge principle is used to derive the Dirac equation and the wave
equation for photons in curved space-times separately $\left[ 10\right] $.
In Section III, we discuss the quantization of the composite spinning
two-particle system. In Section IV, we discuss the solutions of the wave
equation for the photon, which is the analog of the Dirac equation for the
electron. Since this new wave equation is similar to the Dirac equation and
has first order derivatives with respect to the space-time coordinates, it
has a probability interpretation. It has three spin states and describes the
probability amplitudes for the photon. The solutions of this new wave
equation are in the same form as the solutions of the Maxwell equations in
the free space and curved space-time, but the physical meaning of these
functions is different. The solutions to the Maxwell equations and this new
equation represent the amplitude of the energy waves and the probability
waves of the photons, respectively. In Section V, we discuss the radiative
processes and evaluate the scattering amplitudes by using this new
interaction Lagrangian of the photon current and the electron current.
Section VI is the conclusion.

\section{The\ classical\ system}

\noindent \qquad In this section, we develop a formalism to describe the
electron and the photon as the classical spinning particles. Then the phase
space of the classical system is characterized by the pair of conjugate
dynamical variables of the electron and the photon. These are $\left[
x_{1\mu }\left( \tau _{1}\right) ,p_{1\mu }\left( \tau _{1}\right) \right] $
and $\left[ \overline{z}\left( \tau _{1}\right) ,-iz\left( \tau _{1}\right) %
\right] $ for the external and internal dynamical variables of the electron,
respectively, and $\left[ x_{2\mu }\left( \tau _{2}\right) ,p_{2\mu }\left(
\tau _{2}\right) \right] $ and $\left[ \eta ^{\dagger }\left( \tau
_{2}\right) -i\eta \left( \tau _{2}\right) \right] $ for the external and
internal dynamical variables\ of the photon. $x^{\mu }$ and $p^{\mu }$ are
four vectors of the coordinate and momentum, respectively. Here, $\tau _{1}$
and $\tau _{2}$ are the invariant time parameters of the electron and
photon, respectively. $\overline{z}$ and $z$ are four component complex
spinors, which represent internal degrees of freedom for a particle $\left( 
\text{the electron}\right) $ with negative energy. $\eta ^{\dagger }$ and $%
\eta $ are two component complex spinors which represent internal degrees of
freedom for\ a particle $\left( \text{the photon}\right) $ without negative
energy. The units are $c=$ $\hbar =1$.

The action of the free electron and photon system is given in the Cartan
form: 
\begin{equation}
A=\int \left[ p_{1\mu }\left( dx_{1}^{\mu }-\overline{z}\gamma ^{\mu }zd\tau
_{1}\right) -izd\overline{z}+p_{2\mu }\left( dx_{2}^{\mu }-\eta ^{\dagger
}\sigma ^{\mu }\eta d\tau _{2}\right) -i\eta d\eta ^{+}\right]
\end{equation}%
where $\gamma _{\mu }$ and $\sigma _{\mu }=\left( 1,\overset{\rightharpoonup 
}{\sigma }\right) $ are the Dirac and Pauli matrices, respectively. The
Hamiltonian of the system $\left( H\right) $ is in the covariant form and is
the function of the internal and external dynamical variables of the
particles.

The interaction of the electron and photon can be derived by using the
following transformation for $z$ and $\eta $: 
\begin{equation*}
z\rightarrow e^{i\alpha \left( \tau _{1}\right) }z
\end{equation*}%
\begin{equation*}
\overline{z}\rightarrow \overline{z}e^{-i\alpha \left( \tau _{1}\right) }
\end{equation*}%
\begin{equation*}
\eta \rightarrow e^{i\beta \left( \tau _{2}\right) }\eta
\end{equation*}%
\begin{equation}
\eta ^{\dagger }\rightarrow \eta ^{\dagger }e^{-i\beta \left( \tau
_{2}\right) }
\end{equation}%
We substitute them into Eq. (1). The result is 
\begin{equation*}
A=\int d\tau _{1}\left[ p_{1\mu }\left( \dfrac{dx_{1}^{\mu }}{d\tau _{1}}-%
\overline{z}\gamma ^{\mu }z\right) -\dfrac{d\alpha }{d\tau _{1}}-iz\dfrac{d%
\overline{z}}{d\tau _{1}}\right]
\end{equation*}%
\begin{equation}
+\int d\tau _{2}\left[ p_{2\mu }\left( \dfrac{dx_{2}^{\mu }}{d\tau _{2}}%
-\eta ^{\dagger }\sigma ^{\mu }\eta \right) -\dfrac{d\beta }{d\tau _{2}}-i%
\dfrac{d\eta ^{\dagger }}{d\tau _{2}}\eta \right]
\end{equation}%
We choose $\alpha $ and $\beta $ as 
\begin{equation*}
\dfrac{d\alpha }{d\tau _{1}}=\dfrac{e\left( \lambda \right) }{2}\overline{z}%
\gamma ^{\mu }z\int dx_{2}\delta \left[ x_{1}\left( \tau _{1}\right)
-x_{2}\left( \tau _{2}\right) \right] \eta ^{\dagger }\sigma ^{\mu }\eta
\end{equation*}%
\begin{equation}
\dfrac{d\beta }{d\tau _{2}}=\dfrac{e\left( \lambda \right) }{2}\eta
^{\dagger }\sigma ^{\mu }\eta \int dx_{1}\delta \left[ x_{1}\left( \tau
_{1}\right) -x_{1}\left( \tau _{1}\right) \right] \overline{z}\gamma ^{\mu }z
\end{equation}%
Then the interaction term becomes 
\begin{equation}
A_{\text{int}}=e\left( \lambda \right) \int d\tau _{1}\overline{z}\gamma
^{\mu }z\int d\tau _{2}\int dx_{2}\delta \left[ x_{1}\left( \tau _{1}\right)
-x_{2}\left( \tau _{2}\right) \right] \eta ^{\dagger }\sigma ^{\mu }\eta
\end{equation}%
Thus the $H_{\text{int}}$ is derived by\ the gauge symmetry of the classical
spinors. It consists of the interaction between the electron and photon
currents with the help of $\delta $ function by a\ new coupling constant $%
e\left( \lambda \right) $ which will be obtained in terms of the electron
charge $e.$

We substitute the Eq. (5) into Eq. (3). Then the action becomes 
\begin{eqnarray*}
A &=&\int d\tau _{1}\left[ p_{1\mu }\left( \frac{dx_{1}^{\mu }}{d\tau _{1}}-%
\overline{z}\gamma ^{\mu }z\right) -iz\frac{d\overline{z}}{d\tau _{1}}\right]
\\
&&+\int d\tau _{2}\left[ p_{2\mu }\left( \frac{dx_{2}^{\mu }}{d\tau _{2}}%
-\eta ^{\dagger }\sigma ^{\mu }\eta \right) -i\eta \frac{d\eta ^{+}}{d\tau
_{2}}\right]
\end{eqnarray*}%
\begin{equation}
-e\left( \lambda \right) \int \int d\tau _{1}d\tau _{2}\overline{z}\left(
\tau _{1}\right) \gamma ^{\mu }z\left( \tau _{1}\right) \int dx_{2}\eta
^{\dagger }\left( \tau _{2}\right) \sigma _{\mu }\eta \left( \tau
_{2}\right) \delta \left[ x_{1}\left( \tau _{1}\right) -x_{2}\left( \tau
_{2}\right) \right]
\end{equation}

\subsection{Euler-Lagrange and Hamilton equations}

\noindent \qquad The variations of the action in Eq. (5) are calculated and
the Euler-Lagrange equations of the electron and photon system are obtained.
For the electron, these equations are 
\begin{equation*}
\dfrac{dz}{d\tau _{1}}=-i\gamma ^{\mu }\left[ p_{1\mu }-eA_{\mu }\left(
x_{1}\right) \right] z
\end{equation*}%
\begin{equation*}
\dfrac{d\overline{z}}{d\tau _{1}}=-i\overline{z}\gamma ^{\mu }\left[ p_{1\mu
}-eA_{\mu }\left( x_{1}\right) \right]
\end{equation*}%
\begin{equation*}
\dfrac{dx_{1}^{\mu }}{d\tau _{1}}=\overline{z}\gamma ^{\mu }z
\end{equation*}%
\begin{equation}
\dfrac{dp_{1}^{\mu }}{d\tau _{1}}=-e\overline{z}\gamma ^{\nu }zA_{\nu
,}\,^{\mu }
\end{equation}%
where $A_{\mu }\left( x_{1}\right) $ is the four component vector potential
defined as 
\begin{equation}
-eA_{\mu }\left( x_{1}\right) =e\left( \lambda \right) \int d\tau _{2}\int
dx_{2}\eta ^{\dagger }\sigma _{\mu }\eta \delta \left[ x_{1}\left( \tau
_{1}\right) -x_{2}\left( \tau _{2}\right) \right]
\end{equation}%
which represents the field created by the photon current in the electron's
space-time. The Eq. (7) is in the same form as the equations which were
derived in the electron model in Ref. $\left[ 7\right] $. The Euler-Lagrange
equations of the photon are 
\begin{equation*}
\dfrac{d\eta }{d\tau _{2}}=-i\sigma ^{\mu }\left[ p_{2\mu }-e\left( \lambda
\right) B_{\mu }\left( x_{2}\right) \right] \eta
\end{equation*}%
\begin{equation*}
\dfrac{d\eta ^{\dagger }}{d\tau _{2}}=i\eta ^{\dagger }\sigma ^{\mu }\left[
p_{2\mu }-e\left( \lambda \right) B_{\mu }\left( x_{2}\right) \right]
\end{equation*}%
\begin{equation*}
\dfrac{dx_{2}^{\mu }}{d\tau _{2}}=\eta ^{\dagger }\sigma ^{\mu }\eta
\end{equation*}%
\begin{equation}
\dfrac{dp_{2}^{\mu }}{d\tau _{1}}=eB^{\nu ,\mu }\eta ^{\dagger }\sigma _{\nu
}\eta
\end{equation}%
where $B_{\mu }\left( x_{2}\right) $ is the four component vector potential 
\begin{equation}
-eB_{\mu }\left( x_{2}\right) =e\left( \lambda \right) \int d\tau _{1}%
\overline{z}\gamma _{\mu }z\delta \left[ x_{1}\left( \tau _{1}\right)
-x_{2}\left( \tau _{2}\right) \right]
\end{equation}%
which represents the field created by the electron current in the photon's
space-time. The Hamiltonian equations are the same as the Euler-Lagrange
equations.

\section{Quantization}

\noindent \qquad The configuration space of the two-particle system is $%
M^{4}\otimes C^{4}\otimes M^{4}\otimes C^{2}.$ The quantum states of the
system are represented by the composite wave functions 
\begin{equation*}
\Psi =\Psi \left( x_{1},\overline{z};\tau _{1,}x_{2},\eta ^{\dagger };\tau
_{2}\right)
\end{equation*}%
The state function $\Psi $\ satisfies the Schr\"{o}dinger equation for the%
\textit{\ }composite system in the center of mass system of the proper
times. It is 
\begin{equation}
i\left( \dfrac{\partial }{\partial \tau _{1}}+\dfrac{\partial }{\partial
\tau _{2}}\right) \Psi \left( x_{1},\overline{z};\tau _{1,}x_{2},\eta
^{\dagger };\tau _{2}\right) =\widehat{H}\Psi \left( x_{1},\overline{z};\tau
_{1,}x_{2},\eta ^{\dagger };\tau _{2}\right)
\end{equation}%
where $p_{1}^{\mu }$\ and $p_{2}^{\mu }$\ are the momenta of the electron
and photon respectively, the operator $\widehat{H}$\ is given as 
\begin{equation*}
\widehat{H}=\widehat{\overline{z}}\gamma _{\mu }\widehat{z}p_{1}^{\mu }+%
\widehat{\eta }^{\dagger }\sigma _{\mu }\widehat{\eta }p_{2}^{\mu }-e\left(
\lambda \right) \widehat{\overline{z}}\gamma _{\mu }\widehat{z}\widehat{\eta 
}^{\dagger }\sigma _{\mu }\widehat{\eta }
\end{equation*}%
and the momentum operators representing the internal dynamics of the system
are 
\begin{equation*}
\widehat{z}=\frac{\partial }{\partial \overline{z}},\qquad \widehat{\eta }=%
\frac{\partial }{\partial \eta ^{\dagger }}
\end{equation*}%
In order to obtain the spin eigenstates of the composite system we expand
the $\Psi $\ into the power series of $\overline{z}_{\alpha }$\ and $\eta
_{\beta }^{\dagger }$: 
\begin{equation}
\Psi =\overset{\infty }{\underset{n,m=0}{\sum }}\dfrac{1}{n!m!}\left( 
\overline{z}^{_{\alpha _{0}}}\cdot \cdot \cdot \overline{z}^{\alpha
_{n}}\right) \left( \eta ^{\dagger \beta _{0}}\cdot \cdot \cdot \eta
^{\dagger \beta _{m}}\right) \psi _{\alpha _{0}\cdot \cdot \cdot \alpha
_{n},\,\beta _{0}\cdot \cdot \cdot \beta _{m}}\left( x_{1,}\tau
_{1;}x_{2},\tau _{2}\right)
\end{equation}%
In the Schr\"{o}dinger equation, the magnitudes of the spin are conserved.
Then we obtain the most general eigenvalue equation for the spin-$\frac{n}{2}
$\ and spin-$\frac{m}{2}$\ particles separately by equalizing the
coefficients of the $n^{\text{th }}$power of $\overline{z}$\ and $m^{\text{%
th }}$power of $\eta ^{\dagger }$\ on both sides of the Schr\"{o}dinger
equation: 
\begin{equation*}
\{\overline{z}^{\alpha }\left( \gamma ^{\mu }\widehat{p}_{1\mu }\right)
_{\alpha \alpha ^{^{\prime }}}+\eta ^{\dagger \beta }\left( \sigma ^{\mu }%
\widehat{p}_{2\mu }\right) _{\beta \beta ^{^{\prime }}}
\end{equation*}%
\begin{equation*}
+e\left( \lambda \right) \int d\tau _{1}\overline{z}^{\alpha }\left( \gamma
^{\mu }\right) _{\alpha \alpha ^{^{\prime }}}\eta ^{\dagger \beta }\left(
\sigma _{\mu }\right) _{\beta \beta ^{^{\prime }}}\times \delta \left(
x_{1}-x_{2}\right) \}
\end{equation*}%
\begin{equation*}
\times \sum_{j=1}^{n}\delta ^{\alpha ^{^{\prime }}\alpha _{j}}\overline{z}%
^{\alpha _{1}}\cdot \cdot \cdot \overline{z}^{\alpha _{n}}\times
\sum_{k=1}^{m}\delta ^{\beta ^{^{\prime }}\beta _{k}}\eta ^{\dagger \beta
_{1}}\cdot \cdot \cdot \eta ^{\dagger \beta _{m}}\psi _{\alpha _{1}\cdot
\cdot \cdot \alpha _{n},\,\beta _{1}\cdot \cdot \cdot \beta _{m}}
\end{equation*}%
\begin{equation}
=\epsilon \left( \overline{z}^{_{\alpha _{1}}}\cdot \cdot \cdot \overline{z}%
^{\alpha _{n}}\right) \left( \eta ^{\dagger \beta _{1}}\cdot \cdot \cdot
\eta ^{\dagger \beta _{m}}\right) \psi _{\alpha _{1}\cdot \cdot \cdot \alpha
_{n},\,\beta _{1}\cdot \cdot \cdot \beta _{m}}
\end{equation}%
where $\epsilon $\ is the total energy of the particles in their own rest
frame. Eq. (13) gives the Schr\"{o}dinger equation of the system of higher
spin particles. Thus we derive the eigenvalue equation of the spin-$\frac{1}{%
2}$\ and spin-$1$\ particles as 
\begin{equation*}
\overline{z}_{\alpha }\{\left( \gamma ^{\mu }\widehat{p}_{1\mu }\right)
_{\alpha \alpha ^{^{\prime }}}\eta _{\beta _{1}}^{\dagger }\eta _{\beta
_{2}}^{\dagger }+\delta _{\alpha \alpha ^{\prime }}\eta _{\beta }^{\dagger
}\left( \sigma ^{\mu }\widehat{p}_{2\mu }\right) _{\beta \beta ^{^{\prime
}}}\left( \delta _{\beta ^{^{\prime }}\beta _{1}}\eta _{\beta _{2}}^{\dagger
}+\eta _{\beta _{1}}^{\dagger }\delta _{\beta ^{^{\prime }}\beta _{2}}\right)
\end{equation*}%
\begin{equation*}
+e\left( \lambda \right) \int d\tau \left( \gamma ^{\mu }\right) _{\alpha
\alpha ^{^{\prime }}}\eta _{\beta }^{\dagger }\left( \sigma _{\mu }\right)
_{\beta \beta ^{^{\prime }}}\left( \delta _{\beta ^{^{\prime }}\beta
_{1}}\eta _{\beta _{2}}^{\dagger }+\eta _{\beta _{1}}^{\dagger }\delta
_{\beta ^{^{\prime }}\beta _{2}}\right) \delta \left( x_{1}-x_{2}\right)
\}\psi _{\alpha ^{^{\prime }},\,\beta _{1}\beta _{2}}
\end{equation*}%
\begin{equation}
=\epsilon \overline{z}^{_{\alpha }}\eta _{\beta _{1}}^{\dagger }\eta _{\beta
_{2}}^{\dagger }\psi _{\alpha ,\,\beta _{1}\beta _{2}}
\end{equation}%
If we rewrite this equation in the variational form it becomes 
\begin{equation*}
\int dx_{1}dx_{2}\overline{\psi }_{\alpha _{1},\,\beta _{1}\beta _{2}}\{%
\left[ \gamma ^{\mu }\widehat{p}_{1\mu }\left( 1\otimes 1\right) d\tau
_{1}+\left( \sigma ^{\mu }\otimes 1+1\otimes \sigma ^{\mu }\right) \widehat{p%
}_{2\mu }d\tau _{2}\right]
\end{equation*}%
\begin{equation*}
\times e\left( \lambda \right) \gamma ^{\mu }\left( \sigma ^{\mu }\otimes
1+1\otimes \sigma ^{\mu }\right) \delta \left( x_{1}-x_{2}\right) d\tau
_{1}d\tau _{2}\}_{\left( \alpha _{1},\,\beta _{1}\beta _{2}\right) \,\left(
\gamma _{1},\,\sigma _{1}\sigma _{2}\right) }\psi _{\gamma _{1},\,\sigma
_{1}\sigma _{2}}
\end{equation*}%
\begin{equation}
=\int \int dx_{1}dx_{2}\overline{\psi }_{\alpha _{1},\,\beta _{1}\beta
_{2}}\left( i\frac{\partial }{\partial \tau _{1}}+i\frac{\partial }{\partial
\tau _{2}}\right) \left( 1\otimes 1\right) _{\left( \alpha _{1},\,\beta
_{1}\beta _{2}\right) \,\left( \gamma _{1},\,\sigma _{1}\sigma _{2}\right)
}\psi _{\gamma _{1},\,\sigma _{1}\sigma _{2}}d\tau _{1}d\tau _{2}
\end{equation}

\section{Free electron and photon wave functions}

\noindent \qquad For the weakly interacting particles we separate the
composite wave function $\psi $ in the following way: 
\begin{equation}
\psi \left( x_{1};\tau _{1,}x_{2};\tau _{2}\right) =\varphi \left(
x_{1};\tau _{1}\right) \otimes \phi \left( x_{2};\tau _{2}\right)
\end{equation}%
where $\varphi $ and $\phi $ are the electron and photon wave functions,
respectively. We define $\Sigma ^{\mu }$as 
\begin{equation*}
\Sigma ^{\mu }=\sigma ^{\mu }\otimes 1+1\otimes \sigma ^{\mu }
\end{equation*}%
Thus we obtain the following equations with the help of the variations of $%
\overline{\varphi }$ and $\phi ^{\dagger }$: 
\begin{equation}
\left[ \gamma ^{\mu }\left( \widehat{p}_{1\mu }-eA_{\mu }\right) -m\right]
\varphi =0
\end{equation}%
\begin{equation}
\Sigma ^{\mu }\left( \widehat{p}_{2\mu }-eB_{\mu }\right) \phi =0
\end{equation}%
Equations (17) and (18) are the wave equations for the electron and the
massless-spin-$1$ particle (photon) in the slowly varying field of the
others. In this study we consider the space and time coordinates of the
particles as operators and thus normalize the wave functions $\varphi $ and $%
\phi $ in the four dimensional space-time box with volume $VT.$

\subsection{Coupling constant $e\left( \protect\lambda \right) $}

\noindent $\qquad e\left( \lambda \right) $ is the parameter representing
the coupling of the electron and photon in this new formulation of the
electrodynamics. We compare it with the electron charge $e.$ First we do the
dimension analysis in the definition 
\begin{equation*}
\left[ -eA_{\mu }\left( x\right) =e\left( \lambda \right) \overline{\psi }%
\left( x\right) \gamma _{\mu }\psi \left( x\right) \right]
\end{equation*}%
where $e$ is the dimensionless electron charge. The dimension of the
electromagnetic potential is $\left[ \text{length}\right] ^{-1}$. Since the
wave functions are normalized in four-dimensions, the dimensions of the
product is $\left[ \text{length}\right] ^{-4}$ . Thus $e\left( \lambda
\right) $ is found as 
\begin{equation}
e\left( \lambda \right) =-e\left( VT\right) ^{\frac{3}{4}}
\end{equation}

\subsection{Electron wave function}

\noindent \qquad The expressions of the free electron wave functions,
normalized in four-dimensions are 
\begin{equation*}
\varphi \left( x\right) =\frac{1}{\sqrt{VT}}\int \frac{d^{4}p}{\left( 2\pi
\right) ^{4}}\sqrt{\frac{\left| p_{0}\right| +m}{2\left| p_{0}\right| }}
\end{equation*}%
\begin{equation}
\times \left[ \left( 
\begin{array}{c}
1 \\ 
\frac{\overrightarrow{p}\cdot \overrightarrow{\sigma }}{\left| p_{0}\right|
+m}%
\end{array}%
\right) \otimes \chi ^{\left( s\right) }e^{ip\cdot x}\theta \left( \left|
p_{0}\right| \right) +\left( 
\begin{array}{c}
\frac{\overrightarrow{p}\cdot \overrightarrow{\sigma }}{\left| p_{0}\right|
+m} \\ 
1%
\end{array}%
\right) \otimes \chi ^{\left( s\right) }e^{-ip\cdot x}\theta \left( -\left|
p_{0}\right| \right) \right]
\end{equation}%
where $\chi ^{\left( s\right) }$ is the two component Pauli spinor. The
first (second) term in the bracket corresponds to the positive (negative)
energy solutions of the Dirac equations,which correspond to the forward
(backward) moving electron in time.

\subsection{Photon\ wave function}

\noindent \qquad The eigenvalue equation of the free spin-1 particle
(photon) is 
\begin{equation}
\Sigma ^{\mu }\left( \widehat{p}_{2\mu }\right) \phi =0
\end{equation}%
The solutions to Eq. (21) are 
\begin{equation}
\phi ^{\left( +\right) }\left( x_{\parallel },t\right) =\frac{1}{\sqrt{VT}}%
\left( 
\begin{array}{c}
1 \\ 
0 \\ 
0 \\ 
0%
\end{array}%
\right) e^{-ik\left( t-x_{\parallel }\right) }\ \ \ \text{for}\ \omega =k
\end{equation}%
\begin{equation}
\phi ^{\left( -\right) }\left( x_{\parallel },t_{2}\right) =\frac{1}{\sqrt{VT%
}}\left( 
\begin{array}{c}
0 \\ 
0 \\ 
0 \\ 
1%
\end{array}%
\right) e^{+ik\left( t+x_{\parallel }\right) }\ \ \ \text{for}\ \omega =-k
\end{equation}%
\begin{equation}
\phi ^{\left( 0\right) }\left( x_{\parallel },t\right) =\frac{1}{\sqrt{2VT}}%
\left( 
\begin{array}{c}
0 \\ 
1 \\ 
1 \\ 
0%
\end{array}%
\right) e^{+ikx_{\parallel }}\ \ \ \ \text{for }\ \omega =0,k\neq 0
\end{equation}%
\begin{equation}
\phi ^{\left( 0\right) }\left( x_{\parallel },t\right) =\frac{1}{\sqrt{2VT}}%
\left( 
\begin{array}{c}
0 \\ 
1 \\ 
1 \\ 
0%
\end{array}%
\right) \ \ \ \ \text{for }\ \omega =k=0
\end{equation}%
The first two solutions correspond to the transverse polarized wave
functions with positive and negative helicities and they correspond to the
photon which carries energy and momentum. In the most general case, the
transverse polarized wave functions are written as 
\begin{equation}
\phi \left( x\right) =\frac{1}{\sqrt{2VT}}\left[ a_{+}e^{-ik\cdot
x}+a_{-}e^{ik\cdot x}\right]
\end{equation}%
where the first and second terms in the bracket represent the wave functions
with positive and negative helicities, respectively, $\left( 2VT\right)
^{-1/2}$ is the normalization constant for four dimensional normalization \
and $a_{+}$ and $a_{-}$ are the photon spinors. Consequently, $\phi \left(
x\right) $ represents the photon with energy $k,$ momentum $\overset{%
\rightharpoonup }{k}$ and helicity $\pm 1.$ We obtain $U_{F}\left(
x_{2}-x_{1}\right) ,$ the Green's function of the transverse photon
propagating from point $x_{1}$ to $x_{2}:$%
\begin{equation*}
U_{F}\left( x_{2}-x_{1}\right) =-\frac{1}{\sqrt{VT}}\int \frac{d^{4}k}{%
\left( 2\pi \right) ^{4}}\frac{e^{-i\omega \left( t_{2}-t_{1}\right) +i%
\overset{\rightharpoonup }{k}\cdot \left( \overset{\rightharpoonup }{x}_{2}-%
\overset{\rightharpoonup }{x}_{1}\right) }}{2\omega }
\end{equation*}%
\begin{equation}
\times \left( \frac{1}{\omega -\left| \overset{\rightharpoonup }{k}\right| }+%
\frac{1}{\omega +\left| \overset{\rightharpoonup }{k}\right| }\right)
\end{equation}%
This is equivalent to the Green's function of the plane electromagnetic
waves in the Maxwell theory or in quantumelectrodynamics. The normalizations
and dimensions are different. In the Maxwell theory, the Green's function
represents the transition amplitudes for the energy waves, but in here it
represents the transition amplitudes for the probability waves.

The equations (24) and (25) define the zero-energy longitudinal photon. If
the energy of the photon is zero then the momentum of it is zero or not
zero. The Maxwell fields represent the energy waves and for this reason they
do not have zero energy solutions. We interpret these new solutions: The
third solution (Eq. (24)), $\omega =0,\left| \overset{\rightharpoonup }{k}%
\right| \neq 0$ case is considered as the longitudinal state of the photon
which carries momentum without energy. The energy of the photon emitted from
a heavy nucleus can approximately be zero. In this case, the emitting photon
is longitudinally polarized. We derived the propagator for the photon in the
self interaction of the charged particle or in the interaction of the two
charged particles. The transition amplitudes between the longitudinal
polarized states are 
\begin{equation*}
U_{F}^{\prime }\left( x_{2}-x_{1}\right) =
\end{equation*}%
\begin{equation}
-\frac{1}{\sqrt{VT}}\int \frac{d^{3}k}{\left( 2\pi \right) ^{3}}\frac{e^{i%
\overset{\rightharpoonup }{k}\cdot \left( \overset{\rightharpoonup }{x}_{2}-%
\overset{\rightharpoonup }{x}_{1}\right) }}{2}\left( \frac{1}{\omega -\left| 
\overset{\rightharpoonup }{k}\right| }+\frac{1}{\omega +\left| \overset{%
\rightharpoonup }{k}\right| }\right) \delta \left( \omega \right) d\omega
\end{equation}%
This is the expression of the Green's function of the longitudinal polarized
photon in the limit of zero-energy.

The last solution (Eq. (25)), $\omega =k=0$ case is considered to be a
vacuum state which carries neither energy nor momentum. Historically, there
are many studies on the structure of the vacuum, which have not been
finished completely. In this study we interpret the vacuum as a state
function covering all space and time. In the flat space-time the photon
wave-function is constant. This means that the photon has a constant
probability in space-time even if its energy and momentum are zero. Thus the
photon is neither created nor annihilated; only the energy, momentum and the
polarization of the photon have changed. In QED this is formulated as the
creation and annihilation of the photon. The probability amplitudes between
the emission and absorption points $x_{1}$ and $x_{2}$ or the vacuum states
are 
\begin{equation}
N_{F}\left( x_{1}-x_{2}\right) =\frac{1}{VT}
\end{equation}%
The dimension of this propagator is $\left[ L\right] ^{-4}$

\section{Radiative processes}

\noindent \noindent \qquad The S-matrix element, that will be covered in
this study, is constructed upon the interaction of fields, which are
produced by electron and photon currents. The free-incoming waves of the
electron and photon are represented by $\varphi _{i}\left( x\right) $ and $%
\phi _{i}\left( x\right) $ respectively. $\overline{\varphi }_{f}\left(
x^{\prime }\right) $ and $\phi _{f}^{\dagger }\left( x^{\prime }\right) $
show the outgoing free-waves of the electron and photon, respectively. The $%
S-$matrix elements are dimensionless. Finally, from the dimensional analysis
we have obtained a factor that is represented by $q\left( \theta \right) .$
For each emitting and absorbing longitudinal polarized photon at vertices,
we write 
\begin{equation}
q\left( \theta \right) =\sqrt{\frac{2\pi T}{\omega }}
\end{equation}%
For each emitting and absorbing transverse polarized photon at vertices, 
\begin{equation}
q\left( \theta \right) _{i,f}=\sqrt{\frac{T}{\omega _{i,f}}}
\end{equation}%
where the photon is real and, $f$\ and $i$\ subindexes represent the final
and initial states respectively and $\omega $\ is the energy of transverse
photon. We choose the dimension of Feynman propagator as $\left[ L\right]
^{-4}.$ In this case, we obtain 
\begin{equation}
S_{F}\left( x^{\prime }-x\right) =\frac{1}{\left( VT\right) ^{\frac{1}{4}}}%
\int \frac{d^{4}p}{\left( 2\pi \right) ^{4}}\frac{e^{-ip\left( x-x^{\prime
}\right) }}{\left( \NEG{p}-m\right) }
\end{equation}

\subsection{Compton scattering}

\noindent \qquad The Feynman graph of the Compton scattering, which is one
of the radiative processes, is shown in Fig. 1., where $i$ and $f$
subindexes represent the initial and the final states and $p$ and $k$ are
the four momentum of the electron and photon, respectively. $\in $
represents the photon polarization, and $s$ represents the electron spin. We
can explain the Compton scattering as follows:

In the graph on the left, the electron and photon interact with each other
in the space-time $x$ and propagate to the space-time $x^{\prime }$. Since
the photon does not have energy and momentum between the points $x$ and $%
x^{\prime }$, it is longitudinal polarized. The electron and photon again
interact with each other in the final position $x^{\prime }$ and then
propagate to different space-time regions. In the graph on the right, the
interaction positions are exchanged. The contributions of both cases are
equal to each other. The electron and photon have been propagated into the
future in the space-time. There is a second-order interaction in this
process. The $S$-matrix element of the Compton scattering is 
\begin{equation*}
S_{fi}^{CS}=i\int d^{4}x^{\prime }d^{4}x\overline{\varphi }_{f}\left(
x^{\prime }\right) \left( -i\right) e\left( \lambda \right) \gamma _{\mu
}iS_{F}\left( x^{\prime }-x\right) \left( -i\right) e\left( \lambda \right)
\gamma _{\nu }\varphi _{i}\left( x\right)
\end{equation*}%
\begin{equation*}
\times \lbrack \phi _{f}^{\dagger }\left( x^{\prime }\right) \left(
-i\right) q\left( \theta \right) _{f}\Sigma ^{\mu }iN_{F}\left( x^{\prime
}-x\right) \left( -i\right) q\left( \theta \right) _{i}\Sigma ^{\nu }\phi
_{i}\left( x\right)
\end{equation*}%
\begin{equation}
+\phi _{f}^{\dagger }\left( x\right) \left( -i\right) q\left( \theta \right)
_{f}\Sigma ^{\nu }iN_{F}\left( x-x^{\prime }\right) \left( -i\right) q\left(
\theta \right) _{i}\Sigma ^{\mu }\phi _{i}\left( x^{\prime }\right) ]
\end{equation}%
We carry out the integral upon the coordinates and momenta. Since the photon
spinors are normal to each other, we obtain 
\begin{equation*}
S_{fi}^{CS}=\frac{e^{2}}{VT^{3}}\sqrt{\frac{m^{2}}{E_{f}E_{i}}}\frac{1}{%
\sqrt{2\omega _{f}2\omega _{i}}}\left( 2\pi \right) ^{4}\delta ^{4}\left(
p_{f}+k_{f}-p_{i}-k_{i}\right)
\end{equation*}%
\begin{equation*}
\times \overline{u}\left( p_{f},s_{f}\right) [\left( -i\gamma _{\mu }\right)
a_{+}^{\dagger }\Sigma ^{\mu }\frac{i}{\NEG{p}_{i}+\NEG{k}_{i}-m}\left(
-i\gamma _{\nu }\right) \Sigma ^{\nu }a_{+}
\end{equation*}%
\begin{equation}
+\left( -i\gamma _{\mu }\right) \Sigma ^{\mu }a_{-}\frac{i}{\NEG{p}_{i}-\NEG%
{k}_{f}-m}\left( -i\gamma _{\nu }\right) a_{-}^{\dagger }\Sigma ^{\nu
}]u\left( p_{i},s_{i}\right)
\end{equation}%
Eq. (34) corresponds to the Feynman graphs in Fig. 1, where the
conservations of energy and momentum $p_{f}+k_{f}=p_{i}+k_{i}$ have been
satisfied. This condition corresponds to the process of an incident free
electron and photon transforming to a final state of a free electron and
photon. We define the transverse polarization conditions of the photon as
shown below: 
\begin{equation*}
\gamma _{\mu }a_{+}^{\dagger }\Sigma ^{\mu }\equiv \gamma _{\mu }\in
_{f+}^{\mu }=\notin _{f+},\qquad \gamma _{\nu }\Sigma ^{\nu }a_{+}\equiv
\gamma _{\nu }\in _{i+}^{\nu }=\notin _{i+}
\end{equation*}%
\begin{equation}
\gamma _{\nu }a_{-}^{\dagger }\Sigma ^{\nu }\equiv \gamma _{\nu }\in
_{f-}^{\nu }=\notin _{f-},\qquad \gamma _{\mu }\Sigma ^{\mu }a_{-}\equiv
\gamma _{\mu }\in _{i-}^{\mu }=\notin _{i-}
\end{equation}%
Finally, we obtain 
\begin{equation*}
S_{fi}^{CS}=\frac{e^{2}}{VT^{3}}\sqrt{\frac{m^{2}}{E_{f}E_{i}}}\frac{1}{%
\sqrt{2\omega _{f}2\omega _{i}}}\left( 2\pi \right) ^{4}\delta ^{4}\left(
p_{f}+k_{f}-p_{i}-k_{i}\right) \overline{u}\left( p_{f},s_{f}\right)
\end{equation*}%
\begin{equation}
\times \{\left( -i\notin _{f+}\right) \frac{i}{\NEG{p}_{i}+\NEG{k}_{i}-m}%
\left( -i\notin _{i+}\right) +\left( -i\notin _{i-}\right) \frac{i}{\NEG%
{p}_{i}-\NEG{k}_{f}-m}\left( -i\notin _{f-}\right) \}u\left(
p_{i},s_{i}\right)
\end{equation}

\subsection{Bremsstrahlung}

\noindent \qquad In this process radiation is emitted by an electron bound
to a heavy nucleus with the charge $+Ze$ and the process is shown in Fig. 2.
Since the energy-momentum of the photon incoming to the heavy nucleus is
zero and the momentum of the photon, outgoing from the nucleus, is not zero $%
\left( \overset{\rightharpoonup }{k}\neq 0\right) $, we consider the Coulomb
interaction between the electron and nucleus as the current of the
longitudinal polarized photon. The propagator of the longitudinal polarized
photon is given in Eq. (28). The scattering photon is transverse polarized
and the $S$-matrix element of this process is: 
\begin{equation*}
S_{fi}^{B}=i\int d^{4}x^{\prime }d^{4}xd^{4}x_{0}\overline{\varphi }%
_{f-}\left( x^{\prime }\right) \left( -i\right) e\left( \lambda \right)
\gamma _{\mu }iS_{F}\left( x^{\prime }-x\right) \left( -i\right) e\left(
\lambda \right) \gamma _{0}\varphi _{i-}\left( x\right)
\end{equation*}%
\begin{equation*}
\times \lbrack \phi _{f}^{\dagger }\left( x^{\prime }\right) \left(
-i\right) q\left( \theta \right) _{f}\Sigma ^{\mu }iN_{F}\left( x^{\prime
}-x\right) \left( -i\right) q\left( \theta \right) \Sigma ^{0}iU_{F}^{\prime
}\left( x-x_{0}\right)
\end{equation*}%
\begin{equation*}
+\phi _{f}^{\dagger }\left( x\right) \left( -i\right) q\left( \theta \right)
_{f}\Sigma ^{0}iN_{F}\left( x-x^{\prime }\right) \left( -i\right) q\left(
\theta \right) \Sigma ^{\mu }iU_{F}^{\prime }\left( x^{\prime }-x_{0}\right)
]
\end{equation*}%
\begin{equation}
\times \left( -i\right) q\left( \theta \right) \Sigma _{0}iN_{F}\overline{%
\Psi }\left( x_{0}\right) \left( -i\right) Ze\left( \lambda \right) \gamma
_{0}\Psi \left( x_{0}\right)
\end{equation}%
The current of the nucleus is in the form of $\overline{\Psi }\left(
x_{0}\right) \gamma _{0}\Psi \left( x_{0}\right) $ and its wave function is
a positive energy free-plane wave: 
\begin{equation}
\Psi \left( x_{0}\right) =\frac{1}{\sqrt{VT}}e^{-ip_{n}\cdot x_{0}}b\left(
p_{n},s_{n}\right)
\end{equation}%
Thus we obtain S-matrix 
\begin{equation*}
S_{fi}^{B}=\frac{-Ze^{3}}{VT^{3/2}}\left( 2\pi \right) \delta \left(
E_{f-}+\omega _{f}-E_{i-}\right) \sqrt{\frac{m^{2}}{E_{f-}E_{i-}}}\frac{1}{%
\sqrt{2\omega _{f}}}\frac{1}{\left| \overrightarrow{k}_{f}+\overrightarrow{p}%
_{f-}-\overrightarrow{p}_{i-}\right| ^{2}}
\end{equation*}%
\begin{equation}
\times \overline{u}\left( p_{f-},s_{f-}\right) [\left( -i\notin _{f+}\right) 
\frac{i}{\NEG{p}_{f-}+\NEG{k}_{f}-m}\left( -i\gamma _{0}\right) +(-i\gamma
_{0})\frac{i}{\NEG{p}_{i-}-\NEG{k}_{f}-m}\left( -i\notin _{f+}\right)
]u\left( p_{i-},s_{i-}\right)
\end{equation}%
where $E_{i-}=E_{f-}+\omega _{f}$ represents the conservation of the energy.

\subsection{Annihilation of \ electron-positron pair into gamma rays}

\noindent \qquad The Feynman graph of this process is the same as Compton
scattering. But here, the electron-positron pair decays into two photons.
The interaction is in the order of $e^{2}$. The initial and final photons
are transverse polarized and the photon between the interaction points is
longitudinal polarized. By using the similarity between Compton scattering
and pair annihilation we obtain the following substitutions: 
\begin{equation*}
\begin{array}{ccc}
CS & \leftrightarrow & PA \\ 
\in _{f},k_{f} & \leftrightarrow & \in _{f},k_{f} \\ 
\in _{i},k_{i} & \leftrightarrow & \in _{i},-k_{i} \\ 
p_{f},s_{f} & \leftrightarrow & -p_{+},s_{+} \\ 
p_{i},s_{i} & \leftrightarrow & p_{-},s_{-}%
\end{array}%
\end{equation*}%
Then the matrix elements are 
\begin{equation*}
S_{fi}^{PA}=i\int d^{4}x^{\prime }d^{4}x\overline{\varphi }_{+}\left(
x^{\prime }\right) \left( -i\right) e\left( \lambda \right) \gamma _{\mu
}iS_{F}\left( x^{\prime }-x\right) \left( -i\right) e\left( \lambda \right)
\gamma _{\nu }\varphi _{-}\left( x\right)
\end{equation*}%
\begin{equation*}
\times \{\phi _{f}^{\dagger }\left( x^{\prime }\right) \left( -i\right)
q\left( \theta \right) _{f}\Sigma ^{\mu }iN_{F}\left( x^{\prime }-x\right)
\left( -i\right) q\left( \theta \right) _{i}\Sigma ^{\nu }\phi _{i}\left(
x\right)
\end{equation*}%
\begin{equation}
+\phi _{f}^{\dagger }\left( x\right) \left( -i\right) q\left( \theta \right)
_{f}\Sigma ^{\nu }iN_{F}\left( x-x^{\prime }\right) \left( -i\right) q\left(
\theta \right) _{i}\Sigma ^{\mu }\phi _{i}\left( x^{\prime }\right) \}
\end{equation}%
and it is written finally as 
\begin{equation*}
S_{fi}^{PA}=\frac{e^{2}}{VT^{3}}\sqrt{\frac{m^{2}}{E_{+}E_{-}}}\frac{1}{%
\sqrt{2\omega _{f}2\omega _{i}}}\left( 2\pi \right) ^{4}\delta ^{4}\left(
-p_{+}+k_{f}-p_{-}+k_{i}\right) \overline{\nu }\left( p_{+},s_{+}\right)
\end{equation*}%
\begin{equation}
\times \lbrack \left( -i\notin _{f+}\right) \frac{i}{\NEG{p}_{-}-\NEG%
{k}_{i}-m}\left( -i\notin _{i+}\right) +\left( -i\notin _{i-}\right) \frac{i%
}{\NEG{p}_{-}-\NEG{k}_{f}-m}\left( -i\notin _{f-}\right) ]u\left(
p_{-},s_{-}\right)
\end{equation}%
Here we also have the exchange symmetry of the two photons according to
Bose-Einstein statistics. $\delta $ function in Eq. (41) shows the
conservations of energy and momentum.

\subsection{Pair production}

\noindent \qquad When an energetic photon passes from a region around a
heavy nucleus it decays into an electron-positron pair. We explain this
process as follows: The charge of the heavy nucleus is $-Ze$ and its Coulomb
field is described in a similar way to that of the Bremsstrahlung. When the
transverse polarized initial photon with momentum $k_{i}$ enters into this
field, it gives its energy-momentum to the pair of particles and it becomes
longitudinal polarized.

The Feynman graph of this process is similar to that of the Bremsstrahlung.
By comparing these two processes we obtain the following substitution rules: 
\begin{equation*}
\begin{array}{ccc}
B & \leftrightarrow & PP \\ 
\in _{f},k_{f} & \leftrightarrow & \in _{i},-k_{i} \\ 
p_{f-},s_{f-} & \leftrightarrow & p_{+},s_{+} \\ 
p_{i-},s_{i-} & \leftrightarrow & -p_{-},s_{-}%
\end{array}%
\end{equation*}%
In the momentum space we finally obtain $S$-matrix as 
\begin{equation*}
S_{fi}^{PP}=\frac{-Ze^{3}}{VT^{3/2}}\left( 2\pi \right) \delta \left(
E_{+}-\omega _{i}+E_{-}\right) \sqrt{\frac{m^{2}}{E_{+}E_{-}}}\frac{1}{\sqrt{%
2\omega _{i}}}\frac{1}{\left| -\overset{\rightharpoonup }{k}_{i}+\overset{%
\rightharpoonup }{p}_{+}+\overset{\rightharpoonup }{p}_{-}\right| ^{2}}
\end{equation*}%
\begin{equation}
\times \overline{\nu }\left( p_{+},s_{+}\right) \{\left( -i\notin
_{i+}\right) \frac{i}{\NEG{p}_{+}-\NEG{k}_{i}-m}\left( -i\gamma _{0}\right)
+\left( -i\gamma _{0}\right) \frac{i}{-\NEG{p}_{-}+\NEG{k}_{i}-m}\left(
-i\notin _{i+}\right) \}u\left( p_{-},s_{-}\right)
\end{equation}%
where $\delta $ function shows that the photon energy has been transformed
to electron-positron pairs.

\subsection{Positive energy electron-electron scattering}

\noindent \qquad The process represents the scattering of two identical
particles and the Feynman graph is shown in Fig. 3. The interacting
electrons are labeled with subindexes $1$ and $2$, and their initial and
final states are in the form of free-plane waves. In the first Feynman
graph, the incoming positive energy electron-$1$ is interacted with the
incoming \ zero energy-momentum photon at interaction point $x.$ As a result
of this interaction, a transverse polarized photon is emitted, this photon
interacts with electron-$2$ at interaction point $x^{\prime }$ and loses its
energy-momentum. In the second Feynman graph, electron-$1$ and electron-$2$
have been exchanged in the final state according to Fermi-Dirac statistics.
The corresponding S-matrix element of the process of positive energy
electron-electron scattering is 
\begin{equation*}
S_{fi}^{EES}=i\int d^{4}x^{\prime }d^{4}x[\overline{\varphi }_{f2}\left(
x^{\prime }\right) \left( -i\gamma _{\mu }\right) e\left( \lambda \right)
\varphi _{i2}\left( x^{\prime }\right) iN_{F}\left( -i\Sigma _{0}\right)
q\left( \theta \right) iU_{F}\left( x^{\prime }-x\right)
\end{equation*}%
\begin{equation*}
\times \left( -i\Sigma _{0}\right) q\left( \theta \right) N_{F}\overline{%
\varphi }_{f1}\left( x\right) \left( -i\gamma ^{\mu }\right) e\left( \lambda
\right) \phi _{i1}\left( x\right) -\overline{\varphi }_{f1}\left( x^{\prime
}\right) \left( -i\gamma _{\mu }\right) e\left( \lambda \right) \phi
_{i2}\left( x^{\prime }\right)
\end{equation*}%
\begin{equation}
\times iN_{F}\left( -i\Sigma _{0}\right) q\left( \theta \right) iU_{F}\left(
x^{\prime }-x\right) \left( -i\Sigma _{0}\right) q\left( \theta \right) N_{F}%
\overline{\varphi }_{f2}\left( x\right) \left( -i\gamma ^{\mu }\right)
e\left( \lambda \right) \varphi _{i1}\left( x\right) ]
\end{equation}%
and we finally obtain 
\begin{equation*}
S_{fi}^{EES}=\frac{-e^{2}m^{2}}{V^{3/2}T^{3/2}}\frac{1}{\sqrt{%
E_{f2}E_{i2}E_{f1}E_{i1}}}\left( 2\pi \right) ^{4}\delta ^{4}\left(
p_{f2}+p_{f1}-p_{i2}-p_{i1}\right)
\end{equation*}%
\begin{equation*}
\times \lbrack \overline{u}\left( p_{f2},s_{f2}\right) \left( -i\gamma _{\mu
}\right) u\left( p_{i2},s_{i2}\right) \frac{i}{\left( p_{i1}-p_{f1}\right)
^{2}}\overline{u}\left( p_{f1},s_{f1}\right) \left( -i\gamma ^{\mu }\right)
u\left( p_{i1},s_{i1}\right)
\end{equation*}%
\begin{equation}
-\overline{u}\left( p_{f1},s_{f1}\right) \left( -i\gamma _{\mu }\right)
u\left( p_{i2},s_{i2}\right) \frac{i}{\left( p_{i1}-p_{f2}\right) ^{2}}%
\overline{u}\left( p_{f2},s_{f2}\right) \left( -i\gamma ^{\mu }\right)
u\left( p_{i1},s_{i1}\right) ]
\end{equation}

\subsection{Positive energy (forward moving in time) and negative energy
(backward moving in time) electron scattering}

\noindent \qquad The Feynman graph for this process is similar to positive
energy electron-electron scattering. We obtain the following substitution
rules: 
\begin{equation*}
\begin{array}{ccc}
EES & \leftrightarrow & EPS \\ 
p_{i1} & \leftrightarrow & p_{i-} \\ 
p_{f1} & \leftrightarrow & p_{f-} \\ 
p_{i2} & \leftrightarrow & -p_{f+} \\ 
p_{f2} & \leftrightarrow & -p_{i+}%
\end{array}%
\end{equation*}%
We thus write the S-matrix element as 
\begin{equation*}
S_{fi}^{EPS}=-i\int d^{4}x^{\prime }d^{4}x[\overline{\varphi }_{f+}\left(
x^{\prime }\right) \left( -i\gamma _{\mu }\right) e\left( \lambda \right)
\varphi _{i+}\left( x^{\prime }\right) iN_{F}\left( -i\Sigma _{0}\right)
q\left( \theta \right) iU_{F}\left( x^{\prime }-x\right)
\end{equation*}%
\begin{equation*}
\times \left( -i\Sigma _{0}\right) q\left( \theta \right) N_{F}\overline{%
\varphi }_{f-}\left( x\right) \left( -i\gamma ^{\mu }\right) e\left( \lambda
\right) \phi _{i-}\left( x\right) -\overline{\varphi }_{f-}\left( x^{\prime
}\right) \left( -i\gamma _{\mu }\right) e\left( \lambda \right) \phi
_{f+}\left( x^{\prime }\right)
\end{equation*}%
\begin{equation}
\times iN_{F}\left( -i\Sigma _{0}\right) q\left( \theta \right) iU_{F}\left(
x^{\prime }-x\right) \left( -i\Sigma _{0}\right) q\left( \theta \right) N_{F}%
\overline{\varphi }_{i+}\left( x\right) \left( -i\gamma ^{\mu }\right)
e\left( \lambda \right) \varphi _{i-}\left( x\right) ]
\end{equation}%
and it becomes 
\begin{equation*}
S_{fi}^{EPS}=\frac{e^{2}m^{2}}{V^{3/2}T^{3/2}}\frac{1}{\sqrt{%
E_{f+}E_{i+}E_{f-}E_{i-}}}\left( 2\pi \right) ^{4}\delta ^{4}\left(
p_{f+}-p_{i+}+p_{f-}-p_{i-}\right)
\end{equation*}%
\begin{equation*}
\times \lbrack \overline{\nu }\left( p_{i+},s_{i+}\right) \left( -i\gamma
_{\mu }\right) \nu \left( p_{f+},s_{f+}\right) \frac{i}{\left(
p_{i-}-p_{f-}\right) ^{2}}\overline{u}\left( p_{f-},s_{f-}\right) \left(
-i\gamma ^{\mu }\right) u\left( p_{i-},s_{i-}\right)
\end{equation*}%
\begin{equation*}
-\overline{u}\left( p_{f-},s_{f-}\right) \left( -i\gamma _{\mu }\right) \nu
\left( p_{f+},s_{f+}\right) \frac{i}{\left( p_{i-}+p_{i+}\right) ^{2}}%
\overline{\nu }\left( p_{i+},s_{i+}\right) \left( -i\gamma ^{\mu }\right)
u\left( p_{i-},s_{i-}\right) ]
\end{equation*}%
The first term in the bracket represents the scattering of the positive
energy electron with the negative energy electron. We can interpret the
second term as the exchange of the positive energy electrons and negative
energy electrons according to Fermi-Dirac statistics. By using this
interpretation we can also explain the annihilation of the positronium $%
\left[ 12,13\right] .$ Although the electron and positron do not have the
same charge, they are solutions to the same equation.

\subsection{Vacuum polarization}

\noindent \qquad In vacuum polarization a free transverse polarized photon
is transformed at point $x$ into a longitudinal polarized photon with zero
energy-momentum and an electron-positron pair, and they are annihilated at $%
x^{\prime }.$ The Feynman graph is shown in Fig. 4 and the $S$-matrix for
vacuum polarization is 
\begin{equation*}
S_{fi}^{VP}=-\int id^{4}x^{\prime }d^{4}x\left( -i\right) e\left( \lambda
\right) \gamma _{\alpha \beta }^{\mu }iS_{F}\left( x^{\prime }-x\right)
_{\beta \lambda }\left( -i\right) e\left( \lambda \right) \gamma _{\lambda
\theta }^{\nu }iS_{F}\left( x-x^{\prime }\right) _{\theta \alpha }
\end{equation*}%
\begin{equation}
\times \phi _{f}^{\dagger }\left( x^{\prime }\right) \left( -i\Sigma _{\mu
}\right) q\left( \theta \right) _{f}iN_{F}\left( x^{\prime }-x\right) \left(
-i\Sigma _{\nu }\right) q\left( \theta \right) _{i}\phi _{i}\left( x\right)
\end{equation}%
By performing the $x$ and $x^{\prime }$ integrations we obtain the S-matrix
element of the vacuum polarization as 
\begin{equation*}
S_{fi}^{VP}=-i\frac{e^{2}}{T^{3}}\frac{1}{\sqrt{2\omega _{f}2\omega _{i}}}%
\left( 2\pi \right) ^{4}\delta ^{4}\left( k_{i}-k_{f}\right)
\end{equation*}%
\begin{equation}
\times \int \frac{d^{4}p}{\left( 2\pi \right) ^{4}}\left( -i\right) \gamma
_{\alpha \beta }^{\mu }\in _{\mu f}\frac{i}{\left( \NEG{p}-m\right) _{\beta
\lambda }}\left( -i\right) \gamma _{\lambda \theta }^{\nu }\in _{\nu i}\frac{%
i}{\left( \NEG{p}-\NEG{k}_{i}-m\right) _{\theta \alpha }}
\end{equation}%
The closed fermion loop gives the trace of the matrices. As a result, we
obtain 
\begin{equation*}
S_{fi}^{VP}=-i\frac{1}{T^{3}}\frac{1}{\sqrt{2\omega _{f}2\omega _{i}}}\left(
2\pi \right) ^{4}\delta ^{4}\left( k_{i}-k_{f}\right)
\end{equation*}%
\begin{equation}
\times \int \left( -i\right) e^{2}\frac{d^{4}p}{\left( 2\pi \right) ^{4}}%
\frac{Tr\{\notin _{f}\left( \NEG{p}+m\right) \notin _{i}\left( \NEG{p}-\NEG%
{k}_{i}+m\right) \}}{\left( p^{2}-m^{2}-i\varepsilon \right) \left[ \left(
p-k_{i}\right) ^{2}-m^{2}-i\varepsilon \right] }
\end{equation}%
This integral has a logaritmic singularity at large $p$, hence it is
divergent. The general procedure used to evaluate them is dimensional
regularization and renormalization. A convenient way of the regularization
is to evaluate the divergent terms in the $d$-dimension. We show the
divergent term as $\Pi _{d}^{\mu \nu }\left( k_{i}\right) $, which can be
written as 
\begin{equation}
\Pi _{d}^{\mu \nu }\left( k_{i}\right) =\left( g^{\mu \nu
}k_{i}^{2}-k_{i}^{\mu }k_{i}^{\nu }\right) \Pi _{d}\left( k_{i}^{2}\right)
\end{equation}%
where $\Pi _{d}\left( k_{i}^{2}\right) $ is the scalar part of the equation: 
\begin{equation}
\Pi _{d}\left( k_{i}^{2}\right) =\frac{4e^{2}}{\left( 4\pi \right) ^{d/2}}%
\frac{\Gamma \left( 3-d/2\right) }{\left( 2-d/2\right) \Gamma \left(
2\right) }\underset{0}{\int^{1}}dx\frac{2x\left( 1-x\right) }{\left[
m^{2}-k_{i}^{2}x\left( 1-x\right) \right] ^{2-d/2}}
\end{equation}%
Here the singularity for $d=4$ dimensions is separated. Vacuum polarization
is convergent for $d<4$. The physical result can be obtained from the limit $%
d\rightarrow 4$. To obtain this expression we write the scalar vacuum
polarization in Eq. (50) as the sum of two terms, a finite and an infinite
term. The infinite term is a constant corresponding to the value of $\Pi
_{d}\left( k_{i}^{2}\right) $ at $k_{i}^{2}=0$ and the finite term is
obtained by subtracting the integral in Eq. (50) at $k_{i}^{2}=0$. This
gives 
\begin{equation*}
\Pi _{d}\left( k_{i}^{2}\right) =\Pi _{d}\left( 0\right) +[\Pi _{d}\left(
k_{i}^{2}\right) -\Pi _{d}\left( 0\right) ]
\end{equation*}%
\begin{equation}
=\Pi _{d}\left( 0\right) +\overline{\Pi }\left( k_{i}^{2}\right)
\end{equation}%
We obtain the singular part (the infrared divergent) 
\begin{equation}
\Pi _{d}\left( 0\right) =\frac{\alpha }{\pi }\frac{2\Gamma \left(
1+\varepsilon /2\right) }{\varepsilon \left( 4\pi \right) ^{-\varepsilon /2}}%
\overset{1}{\underset{0}{\int dx}}\frac{2x\left( 1-x\right) }{m^{\varepsilon
}}
\end{equation}%
where $\varepsilon =4-d.$ The finite part is obtained by taking the limit $%
\varepsilon \rightarrow 0:$%
\begin{equation}
\overline{\Pi }_{d}\left( k_{i}^{2}\right) \rightarrow \overline{\Pi }\left(
k_{i}^{2}\right) =\frac{2\alpha }{\pi }\overset{1}{\underset{0}{\int }}%
dx\left( 1-x\right) x\log \left[ 1-\frac{k_{i}^{2}}{m^{2}}x\left( 1-x\right) %
\right]
\end{equation}%
Eq. (53) is zero for $k_{i}^{2}=0$ and it gives 
\begin{equation}
\overline{\Pi }\left( k_{i}^{2}\right) _{\left| k_{i}^{2}\right| \ll
m^{2}}\rightarrow \frac{\alpha }{15\pi }\frac{k_{i}^{2}}{m^{2}}
\end{equation}%
for $\left| k_{i}^{2}\right| \ll m^{2}.$ \ Eq. (53) is complex for $%
k_{i}^{2}\geq m^{2}$. We evaluate the contribution of vacuum polarization
for the longitudinal photon with $k_{i}=0$. As shown in the Feynman graph in
Fig. 5, this corresponds to the production and annihilation of the
positronium. In this process, the initial and the final photons have zero
energy and momentum and are longitudinal polarized.

The S-matrix element is 
\begin{equation*}
S_{\underset{k_{i}=0}{fi}}^{VP}=-\frac{1}{2}i\int d^{4}x^{\prime
}d^{4}x\left( -i\right) e\left( \lambda \right) \gamma _{\alpha \beta }^{\mu
}iS_{F}\left( x^{\prime }-x\right) _{\beta \lambda }
\end{equation*}%
\begin{equation}
\times \left( -i\right) e\left( \lambda \right) \gamma _{\lambda \theta
}^{\nu }iS_{F}\left( x-x^{\prime }\right) _{\theta \alpha }g^{\mu \nu
}iN_{F}\left( -i\Sigma _{0}\right) q\left( \theta \right) iN_{F}\left(
x^{\prime }-x\right) \left( -i\Sigma _{0}\right) q\left( \theta \right) N_{F}
\end{equation}%
This integral gives zero: 
\begin{equation}
S_{\underset{k_{i}=0}{fi}}^{VP}=0
\end{equation}%
This shows that $k_{i}=0$ is the vacuum case for the photon, and positronium
can be produced and annihilated in a vacuum, but this term gives no
contribution to vacuum polarization.

\subsection{Electron self-energy}

\noindent \qquad Electron self energy is explained here as an electron that
interacts the longitudinal photon at $x$ by transforming it to the
transverse state with the energy and momentum photon. At $x^{\prime }$ the
process is reversed. The Feynman graph of the process is shown in Fig. 6.

We write the S-matrix element of the process as 
\begin{equation*}
S_{fi}^{ESE}=i\int d^{4}x^{\prime }d^{4}x\overline{\varphi }_{f}\left(
x^{\prime }\right) \left( -i\gamma _{\mu }\right) e\left( \lambda \right)
iS_{F}\left( x^{\prime }-x\right) \left( -i\gamma _{\nu }\right) g^{\mu \nu
}e\left( \lambda \right) \varphi _{i}\left( x\right)
\end{equation*}%
\begin{equation}
\times iN_{F}\left( -i\Sigma _{0}\right) q\left( \theta \right) iU_{F}\left(
x^{\prime }-x\right) \left( -i\Sigma _{0}\right) q\left( \theta \right) N_{F}
\end{equation}%
This integral can be evaluated by the method used in vacuum polarization. As
a result, we obtain 
\begin{equation}
S_{fi}^{ESE}=-i\frac{1}{V^{1/2}T^{3/2}}\sqrt{\frac{m^{2}}{E_{f}E_{i}}}\left(
2\pi \right) ^{4}\delta ^{4}\left( p_{f}-p_{i}\right) \overline{u}\left(
p_{f},s_{f}\right) \Omega \left( p_{i}\right) u\left( p_{i},s_{i}\right)
\end{equation}%
where $\Omega \left( p_{i}\right) $ gives the self energy of the electron: 
\begin{equation}
\Omega \left( p_{i}\right) =-ie^{2}\int \dfrac{d^{4}k}{\left( 2\pi \right)
^{4}}\dfrac{\gamma ^{\mu }\left( m+\NEG{p}_{i}-\NEG{k}\right) \gamma _{\mu }%
}{\left[ m^{2}-\left( p_{i}-k\right) ^{2}-i\epsilon \right] \left(
-k^{2}-i\epsilon \right) }
\end{equation}%
This integral is divergent. We regularize it using dimensional
regularization and perform the integrals in $d$ dimensions. Then we
substitute $4+\varepsilon $ for $d$ and evaluate $\Omega \left( p_{i}\right) 
$ when $\epsilon \rightarrow 0$:%
\begin{equation*}
\Omega _{\varepsilon }\left( p_{i}\right) =\frac{2\alpha }{\left( 4\pi
\right) ^{1-\varepsilon /2}}\frac{1}{m^{\varepsilon }}\frac{\Gamma \left(
1+\varepsilon /2\right) }{\Gamma \left( 2\right) }\overset{1}{\underset{0}{%
\int }}\frac{dz}{z^{\varepsilon /2}}\{\frac{1}{\varepsilon }\left[ 2m-\left(
1-z\right) \NEG{p}_{i}\right]
\end{equation*}%
\begin{equation*}
+\left( 1-z\right) \NEG{p}_{i}\left[ 1+\left( 1-\frac{\varepsilon }{2}%
\right) \log \left( 1-\frac{p_{i}^{2}\left( 1-z\right) }{m^{2}}\right) %
\right]
\end{equation*}%
\begin{equation}
-m\left[ 1+\left( 2-\frac{\varepsilon }{2}\right) \log \left( 1-\frac{%
p_{i}^{2}\left( 1-z\right) }{m^{2}}\right) \right] \}
\end{equation}%
In Eq. (60) the first term in the curly bracket goes to infinity in the
limiting case $\varepsilon \rightarrow 0$ and the second and third terms are
finite in this limit. If $\ \NEG{p}_{i}=m$, the electron is on the mass
shell. Near the mass shell, that is, when $p_{i}^{2}\approx m^{2}$, we
obtain 
\begin{equation}
\Omega _{\varepsilon }\left( p_{i}\right) =\frac{\alpha }{\left( 4\pi
\right) }\{\frac{1}{\varepsilon }\left[ 3m-\left( \NEG{p}_{i}-m\right) %
\right] -\left( \NEG{p}_{i}-m\right) \log \left( \frac{m^{2}-p_{i}^{2}}{m^{2}%
}\right) \}
\end{equation}%
where the last term has logarithmic singularity. $\Omega \left( p_{i}\right) 
$ becomes complex for $p_{i}^{2}>m^{2}.$ This case corresponds to the
existence of the process of virtual electron decaying into electron and
photon.

\subsection{Energy shifts}

\noindent \qquad In this subsection we evaluate the energy shifts for the
bound electron by using the scattering matrix $S_{fi}$. If we consider Fig.
7, the $S$-matrix element is 
\begin{equation*}
S_{fi}=i\int d^{4}yd^{4}x\overline{\varphi }_{d}\left( y\right) \left(
-i\gamma ^{\mu }\right) e\left( \lambda \right) \varphi _{c}\left( y\right) 
\overline{\varphi }_{b}\left( x\right) \left( -i\gamma ^{\nu }\right) g_{\mu
\nu }e\left( \lambda \right) \varphi _{a}\left( x\right)
\end{equation*}%
\begin{equation}
\times iN_{F}\left( -i\Sigma _{0}\right) q\left( \theta \right) \phi _{\beta
}^{\dagger }\left( y\right) \phi _{\alpha }\left( x\right) \left( -i\Sigma
_{0}\right) q\left( \theta \right) N_{F}
\end{equation}%
where $x=\left( t,\overset{\rightharpoonup }{x}\right) $ and $y=\left(
t^{\prime },\overset{\rightharpoonup }{y}\right) $ are the coordinates of
the interaction points. The propagator of the electron and photon are
written in terms of the wave functions: 
\begin{equation*}
U_{F}\left( y-x\right) =\phi _{\beta }^{\dagger }\left( y\right) \phi
_{\alpha }\left( x\right)
\end{equation*}%
\begin{equation*}
S_{F}\left( y-x\right) =\overline{\varphi }_{c}\left( y\right) \varphi
_{b}\left( x\right)
\end{equation*}%
Generally, $\varphi \left( z\right) =\varphi \left( t,\overset{%
\rightharpoonup }{z}\right) $ is the first quantized wave function of the
electron. The Fourier expansion of this wave function in terms of the time
variable gives 
\begin{equation}
\varphi \left( z\right) =\frac{1}{\sqrt{T}}\underset{n}{\sum }\varphi
_{n}\left( \overset{\rightharpoonup }{z}\right) e^{-iE_{n}t}
\end{equation}%
where $1/\sqrt{T}$ is obtained from the normalization of the function. Eq.
(63) is called Coulomb series expansion, and is different from those used in
standard QED and quantum-optics. As a result, the $S$-matrix element is
obtained: 
\begin{equation*}
S_{fi}=-e^{2}\underset{d,c,b,a}{\sum }2\pi \delta \left(
E_{d}-E_{c}+E_{b}-E_{a}\right) \int d^{3}y\overline{\varphi }_{d}\left( 
\overset{\rightharpoonup }{y}\right) \gamma ^{\mu }\varphi _{c}\left( 
\overset{\rightharpoonup }{y}\right)
\end{equation*}%
\begin{equation*}
\times \int d^{3}x\overline{\varphi }_{b}\left( \overset{\rightharpoonup }{x}%
\right) \gamma _{\mu }\varphi _{a}\left( \overset{\rightharpoonup }{x}%
\right) \int \frac{d^{3}k}{\left( 2\pi \right) ^{3}}e^{i\overset{%
\rightharpoonup }{k}\cdot \left( \overset{\rightharpoonup }{y}-\overset{%
\rightharpoonup }{x}\right) }\frac{1}{2k}
\end{equation*}%
\begin{equation}
\times \{P\left( \frac{1}{E_{d}-E_{c}-k}-\frac{1}{E_{d}-E_{c}+k}\right)
-i\pi \left[ \delta \left( E_{d}-E_{c}-k\right) -\delta \left(
E_{d}-E_{c}+k\right) \right] \}
\end{equation}%
where $P$ is the principal value of the integral and it is in the following
form:%
\begin{equation*}
\frac{1}{E_{d}-E_{c}-\left| \overset{\rightharpoonup }{k}\right| }=P\frac{1}{%
E_{d}-E_{c}-k}-i\pi \delta \left( E_{d}-E_{c}-k\right)
\end{equation*}%
The total energy of the system and the $S$-matrix element depend on each
other with the $\delta $- function. If we consider $\delta \left(
E_{d}-E_{c}+E_{b}-E_{a}\right) $, the energy is conserved according to the
relation of

\begin{center}
$%
\begin{array}{cc}
E_{d}-E_{c}=0\text{ ;} & E_{b}-E_{a}=0%
\end{array}%
$
\end{center}

or 
\begin{equation*}
E_{d}-E_{c}=E_{a}-E_{b}
\end{equation*}%
To obtain the energy shift of a certain state $d$, we eliminate $\delta $
function from the $S$-matrix element in Eq. (64), and we obtain the sum of
all $d$ levels. Thus 
\begin{equation*}
\Delta E_{d}=-2\pi e^{2}\underset{b}{\sum }\int d^{3}x\int d^{3}y\int \frac{%
d^{3}k}{\left( 2\pi \right) ^{3}}e^{i\overset{\rightharpoonup }{k}\cdot
\left( \overset{\rightharpoonup }{y}-\overset{\rightharpoonup }{x}\right) }
\end{equation*}%
\begin{equation*}
\times \left\{ \left( -\frac{P}{k^{2}}\right) \right. \overline{\varphi }%
_{d}\left( \overset{\rightharpoonup }{y}\right) \gamma ^{\mu }\varphi
_{d}\left( \overset{\rightharpoonup }{y}\right) \overline{\varphi }%
_{b}\left( \overset{\rightharpoonup }{x}\right) \gamma _{\mu }\varphi
_{b}\left( \overset{\rightharpoonup }{x}\right)
\end{equation*}%
\begin{equation*}
-\left[ \frac{i\pi }{2k}\left[ \delta \left( E_{d}-E_{b}-k\right) -\delta
\left( E_{d}-E_{b}+k\right) \right] +\frac{P}{2k}\left( \frac{1}{%
E_{d}-E_{b}+k}-\frac{1}{E_{d}-E_{b}-k}\right) \right]
\end{equation*}%
\begin{equation}
\times \left. \overline{\varphi }_{d}\left( \overset{\rightharpoonup }{y}%
\right) \gamma ^{\mu }\varphi _{b}\left( \overset{\rightharpoonup }{y}%
\right) \overline{\varphi }_{b}\left( \overset{\rightharpoonup }{x}\right)
\gamma _{\mu }\varphi _{d}\left( \overset{\rightharpoonup }{x}\right)
\right\}
\end{equation}%
As seen in from Eq. (65), the energy shifts are complex. The first term
corresponds to the contribution of vacuum polarization and the third term
gives Lamb-shift. These terms are real. The second term is imaginary and it
gives the spontaneous emission and absorption. For $\delta \left(
E_{d}-E_{b}-k\right) $ function represents the conservation of energy for
the transformation of the photon at level $d$ and this corresponds to the
spontaneous emission. $\delta \left( E_{d}-E_{b}+k\right) $ represents the
energy conservation for the inverse transformation of the photon at level $b$
and this corresponds to the spontaneous absorption.

\section{Conclusion}

\noindent \qquad In this study we discussed the possibility of the
representation of the photons as spinning particles. First we derived the
Euler-Lagrange equations of the photon as a classical spinning particle and
then studied the current-current interaction of it with the electron. In
this formulation the electron and the photon have equal \ status. In the
Dirac equation the electromagnetic potential is replaced by the velocity
field created by the photon current in the space-time of the electron.
Meanwhile, in the photon wave equation there is a similar term, which
represents the field created by the electron current in the space-time of
the photon. These kind of effects are observed when the motion of light in
intense electron fields is investigated $\left[ 11\right] $.

We obtained the transverse and longitudinal states for the photons and
showed that the transverse photon states correspond to the plane wave
solutions of the Maxwell equations. We also discussed the longitudinal
photon states. The longitudinal, zero energy and non zero momentum states
correspond to the static solutions of the Maxwell equations. We showed the
existence of the zero energy and momentum states of the photons which can be
interpreted as the vacuum states of the photon. In this formulation the
vacuum state has zero energy and momentum, instead of the $\hslash \omega /2$
vacuum energy,\ but it has a nonzero probability amplitude.

Finally we studied the radiative processes of the QED by using the solutions
of the Dirac electron and the photon wave equation and showed that these
processes can be described by representing the photon as a particle with
conserved probability amplitudes, without being created and annihilated and
their $S$-matrix elements give the same result.

We also discussed the generalization of the Pauli principle and
indistinguishability of the electron and its antiparticle (positron) which
are the forward and backward moving solutions of the Dirac equation in time.
We have shown that the particles represented by the solutions of the same
equation can be thought of as identical.

\end{document}